\documentclass[%
aip,
amsmath,amssymb,
preprint,%
]{revtex4-1}

\usepackage{graphicx}
\usepackage{dcolumn}
\usepackage{bm}
\usepackage[utf8]{inputenc}
\usepackage[T1]{fontenc}
\usepackage{mathptmx}
\usepackage{etoolbox}
\usepackage{xcolor}
\usepackage{subcaption}
\usepackage{ulem}

\makeatletter
\def\@email#1#2{%
 \endgroup
 \patchcmd{\titleblock@produce}
  {\frontmatter@RRAPformat}
  {\frontmatter@RRAPformat{\produce@RRAP{*#1\href{mailto:#2}{#2}}}\frontmatter@RRAPformat}
  {}{}
}%
\makeatother
\begin{document}



\title[]{Interplay between electronic and phononic energy dissipation channels in the adsorption of CO on Cu(110)}

\author{Carmen A. Tachino}%
\affiliation{ Grupo de Fisicoqu\'imica en Interfaces y Nanoestructuras, Instituto de F\'isica Rosario (IFIR), CONICET-UNR, Bv. 27 de Febrero 210 bis, S2000EKF Rosario, Argentina}%
\affiliation{Facultad de Ciencias Exactas, Ingenier\'ia y Agrimensura, Universidad Nacional de Rosario, Av. Pellegrini 250,
S2000 Rosario, Argentina}%
%
\author{Federico J. Gonzalez}%
\affiliation{ Grupo de Fisicoqu\'imica en Interfaces y Nanoestructuras, Instituto de F\'isica Rosario (IFIR), CONICET-UNR, Bv. 27 de Febrero 210 bis, S2000EKF Rosario, Argentina}%
\author{Alberto S. Muzas}%
\affiliation{Departamento de Qu\'imica F\'isica Aplicada, Universidad Aut\'onoma de Madrid, 28049 Madrid, Spain}%
\author{J. I\~naki Juaristi}%
\affiliation{Departamento de Pol\'imeros y Materiales Avanzados: F\'isica, Qu\'imica y Tecnolog\'ia, Facultad de Qu\'imicas, Universidad del Pa\'is Vasco (UPV/EHU), Apartado 1072, 20080 Donostia-San Sebasti\'an, Spain}%
\affiliation{Centro de F\'isica de Materiales CFM/MPC (CSIC-UPV/EHU), Paseo Manuel de Lardizabal 5, 20018 Donostia-San Sebasti\'an, Spain}%
\affiliation{Donostia International Physics Center DIPC, Paseo Manuel Lardizabal 4, 20018 Donostia-San Sebasti\'an, Spain}%
\author{Maite Alducin}%
\affiliation{Centro de F\'isica de Materiales CFM/MPC (CSIC-UPV/EHU), Paseo Manuel de Lardizabal 5, 20018 Donostia-San Sebasti\'an, Spain}%
\affiliation{Donostia International Physics Center DIPC, Paseo Manuel Lardizabal 4, 20018 Donostia-San Sebasti\'an, Spain}%
\author{H. Fabio Busnengo}%
\affiliation{ Grupo de Fisicoqu\'imica en Interfaces y Nanoestructuras, Instituto de F\'isica Rosario (IFIR), CONICET-UNR, Bv. 27 de Febrero 210 bis, S2000EKF Rosario, Argentina}%
\affiliation{Facultad de Ciencias Exactas, Ingenier\'ia y Agrimensura, Universidad Nacional de Rosario, Av. Pellegrini 250,
S2000 Rosario, Argentina}%
\date{\today}

\begin{abstract}
In this work, we investigate the relative importance of electronic and phononic energy dissipation during the molecular adsorption of CO on Cu(110). 
Initial sticking probabilities as a function of impact energy for CO impinging at normal incidence at a surface temperature of 90 K were computed using classical trajectory simulations. 
To this aim, we use a full-dimensional potential energy surface constructed using an atomistic neural network trained on density functional theory data obtained with the nonlocal vdW-DF2 exchange-correlation functional.  
Two models are compared: one allowing only energy transfer and dissipation from the molecule to lattice vibrations, and the other also incorporating the effect of molecular energy loss due to the excitation of electron-hole pairs, modeled within the local-density friction approximation. 
Our results reveal, firstly, that the molecule mainly transfers energy to lattice vibrations, and this channel determines the adsorption probabilities, with electronic friction playing a minor role. 
Secondly, once the molecule is trapped near the surface (where electronic density is higher), electron-hole pair excitations accelerate energy dissipation, significantly promoting CO thermalization. 
Still, the faster energy dissipation when electron-hole pair excitations are accounted for accelerates the accommodation of the adsorbed molecules in the chemisorption well but does not significantly alter their lateral displacements over the surface.
\end{abstract}

\maketitle

\section{\label{sec:introduction} Introduction}

Carbon monoxide has long served as a benchmark adsorbate for probing the structure, dynamics, and reactivity of metal surfaces, owing to its simple diatomic nature, high stability and its relevance in heterogeneous catalysis. Studies of CO adsorption provide fundamental insights into molecule-surface interactions and serve as reference systems for validating theoretical models. Among metallic surfaces, copper has attracted special attention. In particular, CO/Cu systems have long served as benchmarks for studying molecule-metal interactions and testing theoretical descriptions of molecule-surface energy exchanges\cite{Yates1994,Vollmer2001,Marinica2002,Gajdos2005,Gajdos2005b,Tremblay2012,Gameel2018,Novko2018,Novko2019,Wei2021,Okabayashi2023a,Okabayashi2023b,Gonzalez2023}. Among the various copper surfaces, Cu(110) stands out due to its open and anisotropic atomic arrangement, which gives rise to well-defined adsorption sites and directional diffusion channels. These structural features make it a prototypical surface for analyzing how surface corrugation influences molecular trapping and energy redistribution. Moreover, CO adsorption on Cu(110) is characterized by a moderate chemisorption well ($\approx$ 0.6 eV) and the absence of dissociative channels
\cite{Rodriguez1989,Harendt1985,Christiansen1992,Ahner1996,Ahner1997,Vollmer2001,Demirci2009}, providing an ideal model to disentangle phonon- and electron-mediated energy exchange processes under well-defined conditions.

The adsorption probability of molecules on metal surfaces can often be accurately described within the adiabatic Born-Oppenheimer approximation, which assumes that electrons adiabatically follow the motion of the nuclei, remaining in their ground state. This is often the case for relatively heavy molecules impinging on transition metal surfaces\cite{Alducin2017}. 
However, several experimental observations and theoretical studies have demonstrated that nonadiabatic effects can play a significant role in different contexts (see, e.g., Refs.~\onlinecite{Saalfrank2006,Frischkorn2006} and references therein).   
Specifically, nonadiabatic effects were found to be relevant in: (i) the broadening of the vibrational linewidth of CO stretching mode on Pt(111) and Cu(100) \cite{Ueba1997,Inoue2016,Novko2018,Novko2019};
(ii) adsorption of atomic and molecular species on thin metal films by detection of chemicurrents across Schottky barriers \cite{Gergen2001,Nienhaus2002}; (iii) energy exchange dynamics of vibrationally excited molecules impinging on noble metal surfaces \cite{Wodtke2016,Wagner2017,Geweke2020,Auerbach2021,Box2021}; 
(iv) surface femtochemistry, where energy exchange between photoexcited electron-hole (e-h) pairs and phonons with the adsorbate degrees of freedom provides the driving force for femtosecond laser-induced processes such as photoinduced diffusion, desorption, and molecular reactions \cite{Saalfrank2006,Frischkorn2006,Juaristi2017}.

These nonadiabatic effects originate from the coupling between the adsorbate motion and low-energy electronic excitations on the substrate, which is typically manifested as the generation of e-h pairs \cite{Wodtke2004,Alducin2017,Alducin2020}. These electronic excitations, together
with the energy exchange with lattice vibrations (phonons),  
determine the net molecule-surface energy exchange\cite{Tully1990}, which ultimately controls whether an impinging molecule becomes adsorbed or reflected from the surface.

To describe these effects theoretically, various dynamical approaches have been developed to incorporate nonadiabatic effects in gas-surface interactions. Among them, ab initio molecular dynamics with electronic friction (AIMDEF) --which accounts for nonadiabatic effects through the local density friction approximation (LDFA) \cite{Juaristi2008prl,BlancoRey2014}-- has been successfully applied to a wide range of gas-surface systems. For instance, AIMDEF has been used to investigate the relaxation of hot H atoms formed upon H$_{2}$ dissociation on Pd(100) and the adsorption of N and N$_{2}$ on transition metals such as Ag(111) and Fe(110) \cite{BlancoRey2014,Novko2015,Novko2016a,Novko2016b}. 
Classical trajectory calculations using potential energy surfaces parameterized from ab initio total energies have also been combined with LDFA under the generalized Langevin oscillator (GLO) model to analyze nonadiabatic energy exchange in systems such as N$_{2}$/W(110), N$_{2}$/W(100) and N/Ag(111) \cite{Goikoetxea2009,Tremblay2010,Martin-Gondre2012,BlancoRey2014}. Subsequent developments extended these studies to laser-induced processes by coupling AIMDEF with electronic and lattice thermal baths within the two-temperature model (2TM)\cite{Anisimov1974}. The electronic-temperature version, ($T_{e}$)-AIMDEF, has been applied to the desorption of H$_{2}$, D$_{2}$ and HD from a H:D-saturated Ru(0001) \cite{Juaristi2017}, while the two-temperature extension, ($T_{e}$, $T_{I}$)-AIMDEF, has been employed to model the laser-induced desorption of O$_2$ from Ag(110)\cite{Loncaric2016a,Loncaric2016b}, CO from Pd(111) \cite{Alducin2019,SerranoJimenez2021,Muzas2022,Muzas2024}, and photoinduced CO desorption and oxidation on (O,CO)/Ru(0001) \cite{Tetenoire2022,Tetenoire2023a,Tetenoire2023b,Zugec2024}. 
Moreover, machine learning interatomic potentials (MLIPs) have recently proven to be a powerful route for exploring nonadiabatic dynamics, capturing the full dimensionality of the surface atom movements at a fraction of the cost of direct ab initio approaches\cite{Zhang2020,SerranoJimenez2021,Muzas2022,Lindner2023,Muzas2024,Zugec2024,Mladineo2025,Stark2025,Wang2025}. In this context, a recent study used an MLIP to describe the potential energy surface (PES) combined with LDFA to study femtosecond laser-induced CO diffusion and desorption on Cu(110), providing further insight into the interplay between electronic and phononic energy exchange channels under laser irradiation on a weak electron–phonon coupling surface like Cu\cite{Gonzalez2025}. 

Determining the relative importance of phononic and electronic channels remains a central issue. 
Several studies on adsorption dynamics have shown that the relevance of nonadiabatic effects varies from system to system, depending on factors such as the mass of the impinging atom or molecule relative to the surface atoms, equilibrium distances, and other specific variables\cite{Juaristi2008prl,Goikoetxea2012,BlancoRey2014,Novko2015,Novko2016a,Novko2016b,Alducin2017,Loncaric2019,Stark2025,Preston2025}. The relative role of phononic and electronic channels remains an open question, particularly under conditions where both mechanisms coexist and compete. 

Motivated by this open question, we investigate the interplay between these two channels for CO adsorption on Cu(110) by employing a classical molecular dynamics approach that incorporates both phononic and electronic energy exchange mechanisms. Within this framework, the effect of e-h pair excitations is represented through the LDFA. Quasi-classical trajectories (QCT) simulations are performed on a full-dimensional artificial neural network potential energy surface (ANN-PES), which was previously used to study the dynamics of CO adsorption on Cu(110) accounting only for molecular energy transfer to lattice vibrations \cite{Gonzalez2023}. 
Here, we extend that analysis to quantify the relative contributions of phononic and electronic channels in the molecular adsorption of CO on Cu(110).

\section{Theoretical Methodology}\label{sec:theory}

\subsection{DFT calculations and ANN-PES}\label{subsec:DFT-ANN}
All dynamics simulations presented in this work were performed using the ANN-PES previously developed and validated for the CO/Cu(110) system \cite{Gonzalez2023}. The methodology used for the generation of the DFT dataset and the training of the PES has been described in detail in Ref. \onlinecite{Gonzalez2023}.  

The DFT database used to train the ANN-PES was generated with the Vienna ab initio simulation package (VASP) \cite{Kresse1993a,Kresse1994a,Kresse1994b,Kresse1996a,Kresse1996b,Kresse1999} employing the nonlocal vdW-DF2 exchange-correlation functional \cite{Lee2010} to account for dispersion interactions relevant to molecule-metal systems. The projector augmented-wave (PAW) method \cite{Blochl1994} was used to describe the electron-ion interaction, with a plane-wave energy cutoff of 400 eV. The Cu(110) surface was modeled as a five-layer slab in a (3$\times$2) periodic supercell with an interlayer vacuum spacing between consecutive slabs of 13.55 {\AA}. The two bottom layers of the slab were kept fixed, while the three topmost layers were allowed to move. Brillouin zone sampling was carried out using a 7$\times$7$\times$1 \textbf{k}-point grid within the Monkhorst-Pack scheme \cite{Monkhorst1976}.

The ANN-PES was constructed using the {\ae}net code \cite{Artrith2016}, following the Behler–Parrinello high-dimensional neural network formalism \cite{Behler2007}. The total energy is expressed as a sum of atomic energy contributions, each predicted by a feed-forward neural network with two hidden layers, consisting of ten neurons per hidden layer and hyperbolic tangent activation functions. Radial and angular symmetry functions were employed to represent the local chemical environments of C, O, and Cu, using optimized parameters as described in Ref.~\onlinecite{Gonzalez2023}. The network was trained to minimize the root-mean-square error (RMSE) between DFT and ANN energies over 7\,803 configurations. 

The resulting ANN-PES reproduces the DFT potential with a RMSE of 0.03 eV ($\approx$ 0.9 meV/atom) and yields the adsorption energy ($\approx$0.6 eV) and energy barriers for CO diffusion in excellent agreement with DFT values \cite{Gonzalez2023}. This level of accuracy ensures a reliable representation of the multidimensional energy landscape required for large-scale molecular-dynamics simulations. 

\subsection{QCT simulation details}\label{subsec:friction}
To investigate the role of electronic effects on CO adsorption over Cu(110), we performed simulations using two models: one considering energy dissipation exclusively into lattice vibrations (hereafter referred to as Ph), and another including molecular energy transfer and dissipation due to both phonon and e-h pair excitations (hereafter referred to as Ph+EF).

In the simulations considering only energy exchange with phonons, the interaction between the adsorbate and the lattice vibrations is naturally incorporated through the dependence of the multidimensional ANN-PES ,$V\!\left( \{{\bf r}_{j}\} \right)$, where $\{{\bf r}_{j}\}$ represents the ensemble of all nuclear coordinates of the system.  
The motion of the molecular atoms (C and O) and the Cu atoms in the two topmost layers, indicated with the subindex $i$, is obtained by integrating the classical equations of motion:

\begin{equation}
	\label{eq:onlyPhonos} m_{i}\,{\bf\ddot{r}}_{i}=-\nabla_{i}V\left( \{{\bf r}_{j}\} \right) \,.
\end{equation}
To ensure the thermalization of the surface at $T_{ph}$ (90 K in our simulations) and to compensate for the finite size of the simulation cell, the dynamics of the mobile Cu atoms in the third layer are coupled to a Nos\'e–Hoover thermostat \cite{Nose1984,Nose1991,Hoover1985}. This heat bath maintains the target temperature and prevents unrealistic heating of the slab that would otherwise result from the energy transferred by the impinging molecule. The time evolution of the $k$-th Cu atom in this thermostatted layer is governed by:

\begin{equation}
    \label{eq:Nose1} m_{Cu}\,{\bf\ddot{r}}_{k}=-\nabla_{k}V\left( \{{\bf r}_{j}\} \right) - \zeta \,m_{Cu}\, {\bf\dot{r}}_{k},
\end{equation}
where the evolution of the thermostat variable $\zeta$ is given by:

\begin{equation}
    \label{eq:Nose2} \dot{\zeta}=\frac{1}{Q}\, \left[ \sum_{j=1}^{N_{Cu}} m_{Cu}\, \left| {{\bf\dot{r}}_{j}} \right|^{2} - 3\,N_{Cu}\,k_{B}\,T_{ph} \right],
\end{equation}
where $Q$ is the Nos\'e-Hoover mass parameter \cite{Frenkel2002}, set in the present work to $Q$=250 eV fs$^{2}$. 

In the simulations that also incorporate nonadiabatic effects, the dynamics of the adsorbate atoms (C and O) are modeled using a Langevin-type equation to account for energy loss due to e-h pair excitations:

\begin{equation}
	\label{eq:friction} m_{i}\,{\bf\ddot{r}}_{i}=-\nabla_{i}V\left( \{{\bf r}_{j}\} \right) - \eta_{i}\left( {\bf r}_{i} \right)\, {\bf\dot{r}}_{i}+{\bf G}_{e}[T_{e},\eta_{i}({\bf r}_{i})],
\end{equation}
where the second and third terms on the r.h.s. correspond to the electronic friction and the random force, respectively. These terms describe the coupling of the adsorbate motion to the electronic bath of the metal, characterized by a constant temperature $T_{e}$ (90 K in our simulations) \cite{Alducin2020}.

The friction coefficient $\eta_{i}\left( {\bf r}_{i}\right)$ 
is evaluated within the LDFA \cite{Juaristi2008prl}, which has been found suitable to describe the effects of e-h pair excitations for numerous molecule-surface systems\cite{Gonzalez2025,Novko2016b,Alducin2017}. 
This coefficient is determined by the local electronic density of the clean metal surface, $n_{surf}({\bf r}_{i})$, at the position of the corresponding atom. Additional details regarding the parametrization for C and O atoms and the calculation of $n_{surf}({\bf r}_{i})$ at each step of the simulation are provided in Ref.~\onlinecite{Gonzalez2025}. 

The random force, ${\bf G}_{e}[T_{e},\eta_{i}({\bf r}_{i})]$, can be modeled according to the second fluctuation– dissipation theorem by a Gaussian white noise with variance 

\begin{equation}
    \label{eq:variance} Var[{\bf G}_{e}(T_{e},\eta_{i}({\bf r}_{i}))] = \frac{2\,k_{B}\,T_{e}\,\eta_{i}({\bf r}_{i})}{\Delta t},
\end{equation}
where $k_{B}$ is the Boltzmann constant and $\Delta t$ is the integration time.

This framework was used to perform QCT simulations, using a time step of 0.5 fs.  
The initial positions of the molecular center of mass were placed 8.5 {\AA} above the surface. The lateral coordinates were uniformly distributed across the simulation cell area, whereas the initial molecular orientations were randomly sampled to ensure a uniform distribution over the full solid angle of the C-O internuclear vector. 
The molecules were initially non-rotating, with the initial internuclear distances and velocities assigned using the procedure described in Ref.~\onlinecite{Gonzalez2023}, and consistent with the vibrational zero-point energy (ZPE) of CO in vacuum, $\text{ZPE}=0.13$ eV. 
For each incidence energy, 3000 trajectories were computed with a maximum integration time of 20 ps ($t_f=20$ ps).  
A trajectory is classified as \textit{reflected} if, at some time instant during the simulation, the Z coordinate of the molecular center-of-mass (Z$_\text{CM}$) exceeds its initial value and the momentum component normal to the surface is pointing toward the vacuum; otherwise, it is considered \textit{adsorbed}.

We verified that simulations incorporating only energy exchange with phonons are consistent with those values reported in Ref.~\onlinecite{Gonzalez2023}, which used the same ANN-PES but employed a different implementation for the thermostat, obtaining discrepancies for each value of the initial sticking probability (S$_0$) below 2$\%$.

\section{\label{sec:res}Results and Discussion}

\subsection{
Relevance of e-h pair excitations for CO molecular adsorption}
\label{sec:S0}

Molecular adsorption is inherently a dissipative process; for a molecule to stabilize on the surface, it must transfer its energy to the substrate via coupling to lattice vibrations (phonons) or electronic excitations (e-h pairs). Therefore, to accurately compute molecular adsorption probabilities and describe the dynamics of this process, at least one of these energy dissipation mechanisms must be included in the modeling.  
Calculations performed within the Born-Oppenheimer static surface (BOSS) approximation --which omit all non-adiabatic and phononic energy exchange channels between the molecule and the surface-- yield S$_0 \approx 0$ for any initial impact energy (see below), provided that a sufficiently long integration time is considered.

\begin{figure}[htb!]
\includegraphics[width=0.6\linewidth]{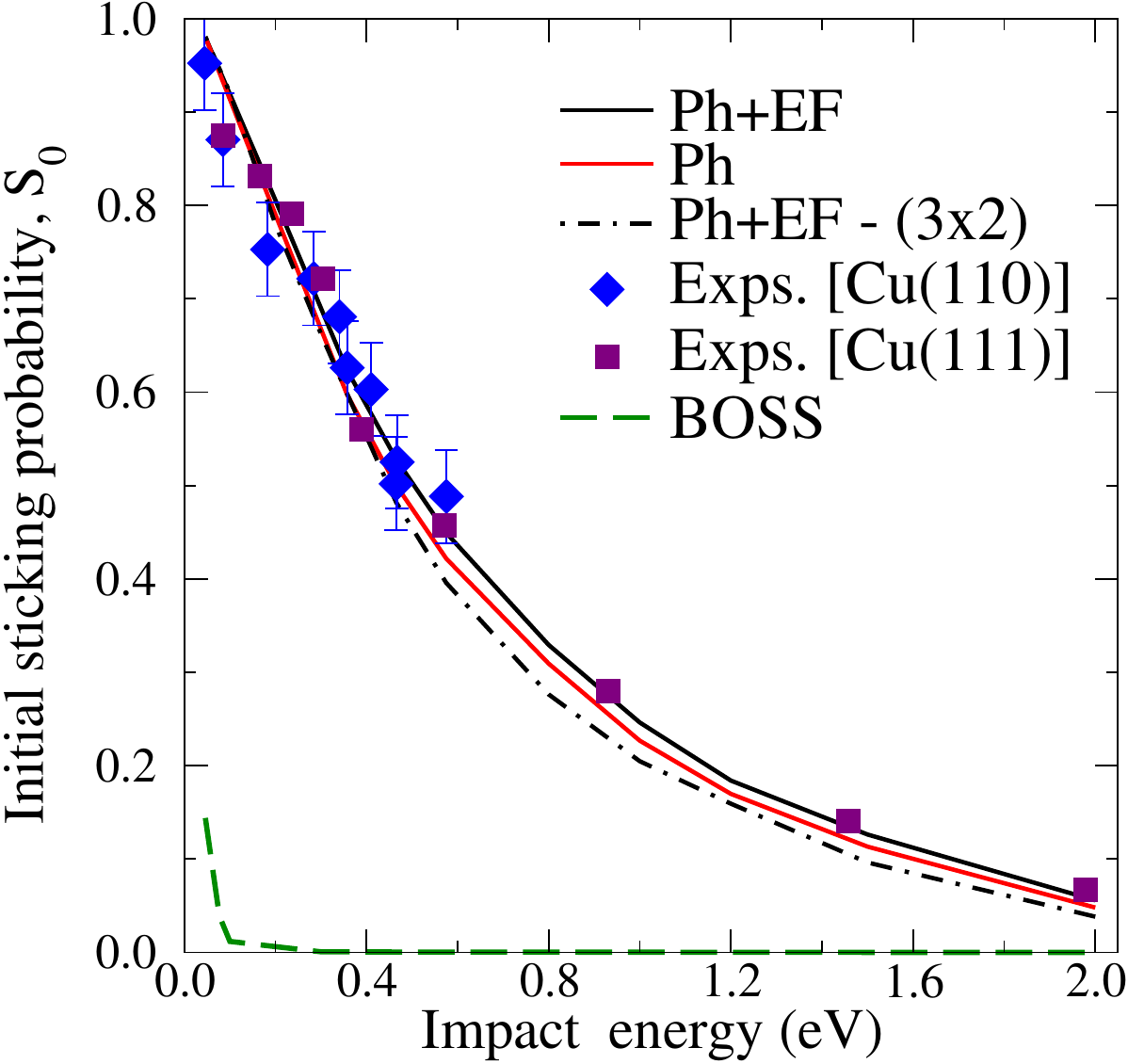}
\caption{S$_{0}$ values as a function of the initial CM translational kinetic energy of the molecule, E$_{\mathrm{i}}$. 
Black (red) solid line, Ph+EF (Ph) for (6$\times$5) supercell; black dot-dashed line, Ph+EF for (3$\times$2) supercell; green dashed line, BOSS model for (6$\times$5) supercell. 
The blue diamonds represent experimental data for CO/Cu(110) at T$_{\mathrm{S}}$=90 K \cite{Kunat2001}, and violet squares correspond to experiments for CO/Cu(111) at T$_{\mathrm{S}}$=85 K \cite{Kneitz1999a,Kneitz1999b}.}  
    \label{fig:S0_6x5}
\end{figure}

To evaluate the relative roles of the two mentioned energy exchange and dissipation mechanisms, Fig.~\ref{fig:S0_6x5} shows the S$_0$(E$_i$) curves obtained from Ph and Ph+EF calculations (these acronyms and their physical interpretations were defined in Sec.~\ref{subsec:friction}). Results from the BOSS model are also included to illustrate the necessity of accounting for energy transfer effects when calculating molecular adsorption probabilities.   
The obtained theoretical results are compared to available experimental results using supersonic molecular beams (SMBs). Note that although the experimental results, represented by violet squares, correspond to Cu(111), they are in perfect agreement with the results for Cu(110) across the entire energy range for which data exists for both surfaces (i.e., $0<\text{E}_i<0.6$ eV). Therefore, comparing our results with the Cu(111) experiments from Refs.~\onlinecite{Kneitz1999a,Kneitz1999b} is valid and useful, as it allows us to extend the energy range considered up to E$_i=2$ eV.

The Ph and Ph+EF calculations yield S$_0$(E$_i$) curves that are very similar both qualitatively and quantitatively, and they are in perfect agreement with the experimental data across the entire energy range considered. The similarity of both theoretical sticking curves indicates that the relative role of the molecule-surface energy exchange due to the excitation of e-h pairs is small. In other words, energy exchange between the molecule and the surface vibrations (phonons) appears to be the main mechanism responsible for the molecular adsorption of CO on Cu(110).

To corroborate this interpretation, we also performed Ph+EF calculations for a smaller (3$\times$2) simulation supercell. Note that performing Ph+EF calculations for different supercell sizes, in principle, only appreciably modifies the manner in which the energy exchange between the molecule and the surface vibrations is described, since the cell size does not alter the electronic density generated by the surface at the positions of the molecule during the trajectories. Conversely, varying the supercell size from (6$\times$5) to (3$\times$2) significantly modifies how the model allows for the dissipation of the energy transferred by the molecule to the surface. In this sense, larger cells reduce the spurious effect of reflection of the transferred energy at the supercell faces due to the use of periodic boundary conditions.

Note that the difference between the S$_0$(E$_i$) curves (for Ph+EF simulations) for the (6$\times$5) and (3$\times$2) supercells is larger than the difference between the curves obtained for the (6$\times$5) cell in Ph and Ph+EF simulations. This is in line with the previous interpretation that the energy exchange and dissipation mechanism of the molecule to the surface vibrations (phonons) is the primary factor responsible for the molecular adsorption of CO on Cu(110).

Molecular adsorption onto a surface below the desorption temperature of the adsorbate ($\sim$200 K in the case of CO/Cu(110)\cite{Kunat2001}) is a process that starts when the molecule approaches the surface and begins to interact with it, and it can be considered to end when the adsorbate reaches thermal equilibrium on the surface. Thermalization can take a very long time depending on the exothermicity of molecular adsorption and the efficacy of the molecule-surface energy exchange and dissipation mechanisms. However, this does not mean that the dynamics along this long time interval is important in determining S$_0$.

For a better understanding of the chemisorption process, its causes and its effects, one can separate the total time interval of the molecular adsorption process into two parts: one during which the dynamics determines the final value of S$_0$ (hereafter referred to as the adsorption time interval), and a forthcoming one during which the molecules that will eventually chemisorb simply accommodate on the surface until thermalization is complete (hereafter referred to as the accommodation time interval). 

Note that non-adiabatic effects evidenced by experiments during chemisorption (e.g., chemicurrents) originate, in principle, during the entire time interval that the chemisorption process takes.

However, the existence of such non-adiabatic effects does not mean that they are determinant of the adsorption probability. Actually, if most of the molecular energy is transferred and dissipated to the electron bath (through e-h pair excitations) during the accommodation period, such non-adiabatic effects should barely influence the measured values of S$_0$. In contrast, if the adsorption period is long, allowing the molecule to remain close to the surface before the dynamics determines whether it will be chemisorbed or reflected back to vacuum, the excitation of e-h pairs might significantly affect the adsorption probability.

In order to shed light on the duration of the adsorption time interval relative to the whole chemisorption process, we have computed the fraction of molecules that remain close to the surface, F$_\text{close}$, as a function of time. Here, we consider that a molecule remains close to the surface if the Z coordinate of its center-of-mass (Z$_{cm}$) is closer to the surface than in the initial state, i.e., Z$_{cm}<8.5$ \AA.
 
\begin{figure}[htb!]
	\centering
	\includegraphics[width=0.6\linewidth]{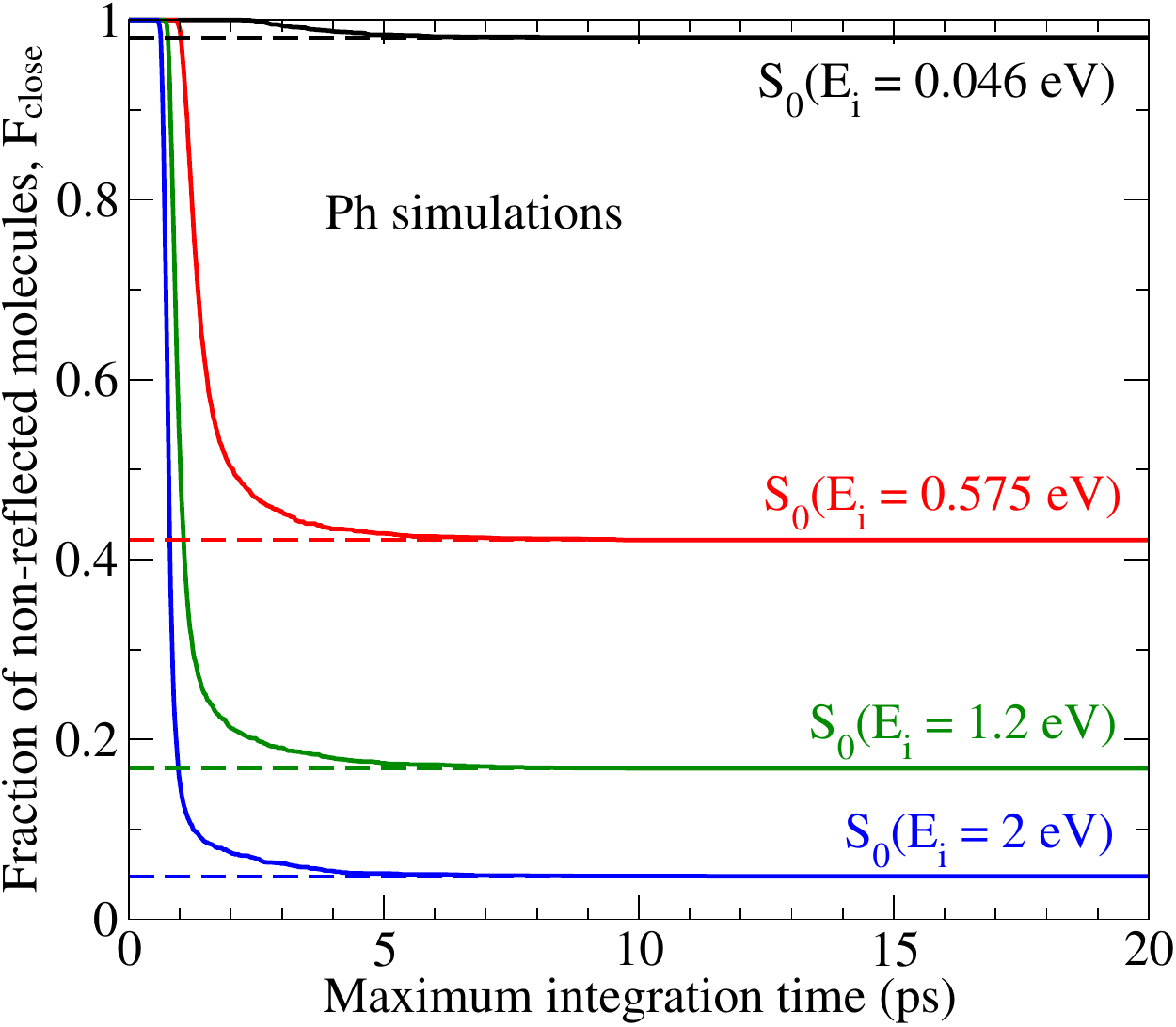}  
	\caption{Fraction of molecules that remain close to the surface, F$_\text{close}$, as a function of maximum integration time for Ph simulations at different incidence energies: 0.046, 0.575, 1.2, and 2 eV. Asymptotic S$_0$ values are represented by dashed lines.}
    \label{fig:reflection_times}
\end{figure}

The results obtained in Ph simulations are shown in Fig.~\ref{fig:reflection_times}. Obviously, since we simulate a SMB experiment in which all the molecules are initially far and directed towards the surface, F$_\text{close}=1$ for small values of $t$. At the time at which some molecules start reaching the initial value of the molecule-surface distance after interacting with the surface, F$_\text{close}$ decreases, reaching the asymptotic value of F$_\text{close}$ for long times
equal to S$_0$. Figure~\ref{fig:reflection_times} can then be used to estimate an approximate value of the duration of what we have called the adsorption time interval. 

It is observed that for all the initial impact energies considered in Fig.~\ref{fig:reflection_times}, F$_\text{close}$ at 4-5 ps is slightly greater but approximately equal S$_0$. Exactly the same conclusion is reached if, instead of plotting F$_\text{close}$(t) for Ph simulations, we plot F$_\text{close}$(t) for Ph+EF simulations. 
Quantitatively, the F$_\text{close}$ values at 5 ps differ by less than 0.01 with respect to the asymptotic S$_{0}$ values. 
Thus, while for the CO/Cu(110) system and the energy range considered in this work, the duration of the adsorption time interval is between 4 and 5 ps, for simplicity in what follows, we will adopt the value of 5 ps.

The overall insensitivity of S$_{0}$ to the incorporation of electronic friction effects can be attributed to the efficient energy transfer to lattice vibrations, which is strongly favored by the relatively large mass ratio between the CO adsorbate and the surface atoms ($\gamma=m_{CO}/m_{Cu} \approx$ 0.44). 
Such a ratio facilitates momentum transfer to phonons rather than to the electronic channel. Similar conclusions were reported in previous AIMDEF studies for relatively heavy adsorbates interacting with transition metal surfaces \cite{Martin-Gondre2012,Novko2015,Novko2016a,Novko2016b,Alducin2017}. In the following subsection, we analyze in detail the energy exchange between the molecule and the vibrations of the Cu atoms, as well as the energy that is dissipated due to e-h pair excitations.

\subsection{Time-resolved energy dissipation}
\label{sec:average}

\begin{figure}[htb!]
	\centering
	\includegraphics[width=1.0\linewidth]{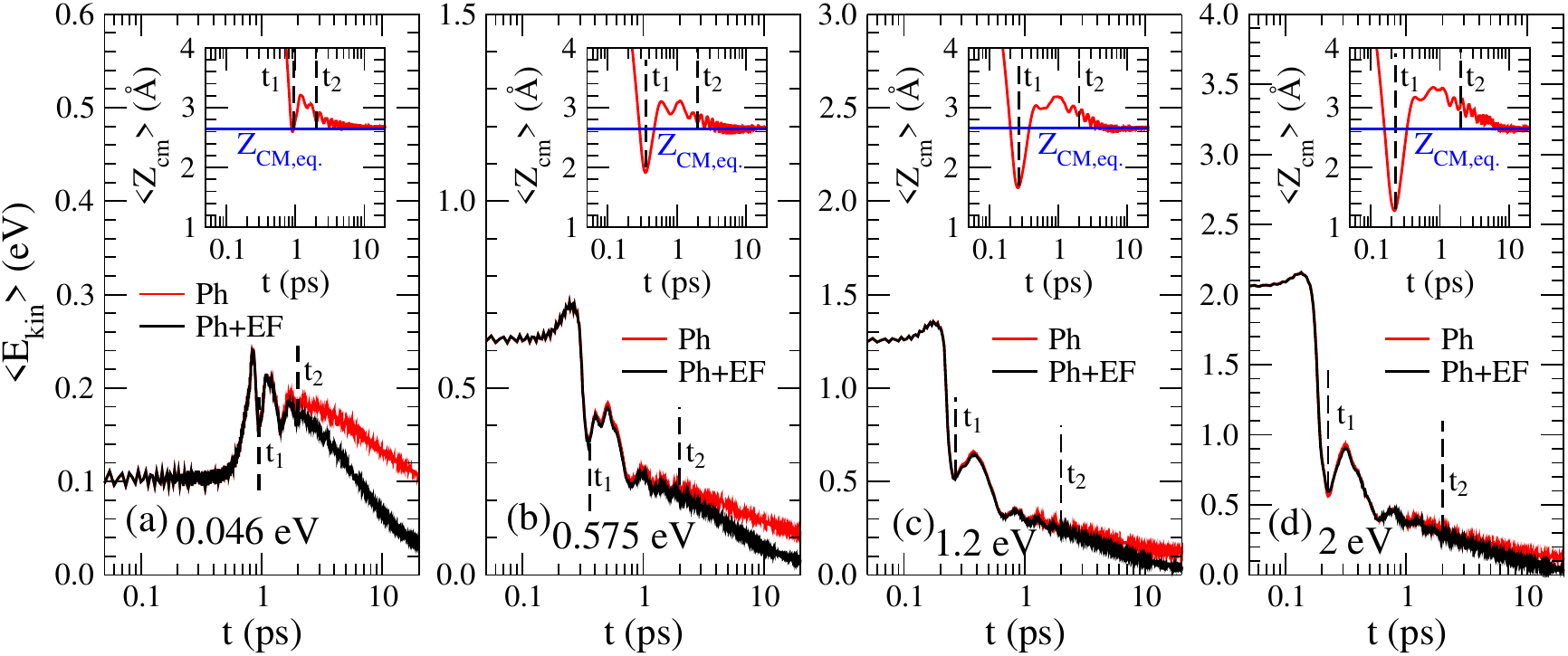}  
	\caption{Average total kinetic energy of adsorbed CO molecules, $\left\langle E_{kin} \right\rangle $ as a function of time, for different impact energies: (a) 0.046 eV, (b) 0.575 eV, (c) 1.2 eV, and (d) 2.0 eV. Insets show the average molecular CM, $\left\langle Z_{CM}\right\rangle$ as a function of time for Ph simulations (Ph+EF results are similar and have been omitted for clarity). Red: including energy exchange with phonons, black: including phonons and electronic friction.}
    \label{fig:Ekintot_6x5}
\end{figure}

Figure~\ref{fig:Ekintot_6x5} displays the time evolution of the mean total kinetic energy, $\left\langle E_{kin} \right\rangle $, for molecules classified as adsorbed. This quantity is defined as the ensemble average over the set of trajectories classified as adsorbed.
The initial values are equal to $ \left\langle E_{kin}(t=0) \right\rangle=E_i+\text{ZPE}/2$.

\begin{figure}[htb!]
	\includegraphics[width=0.8\linewidth]{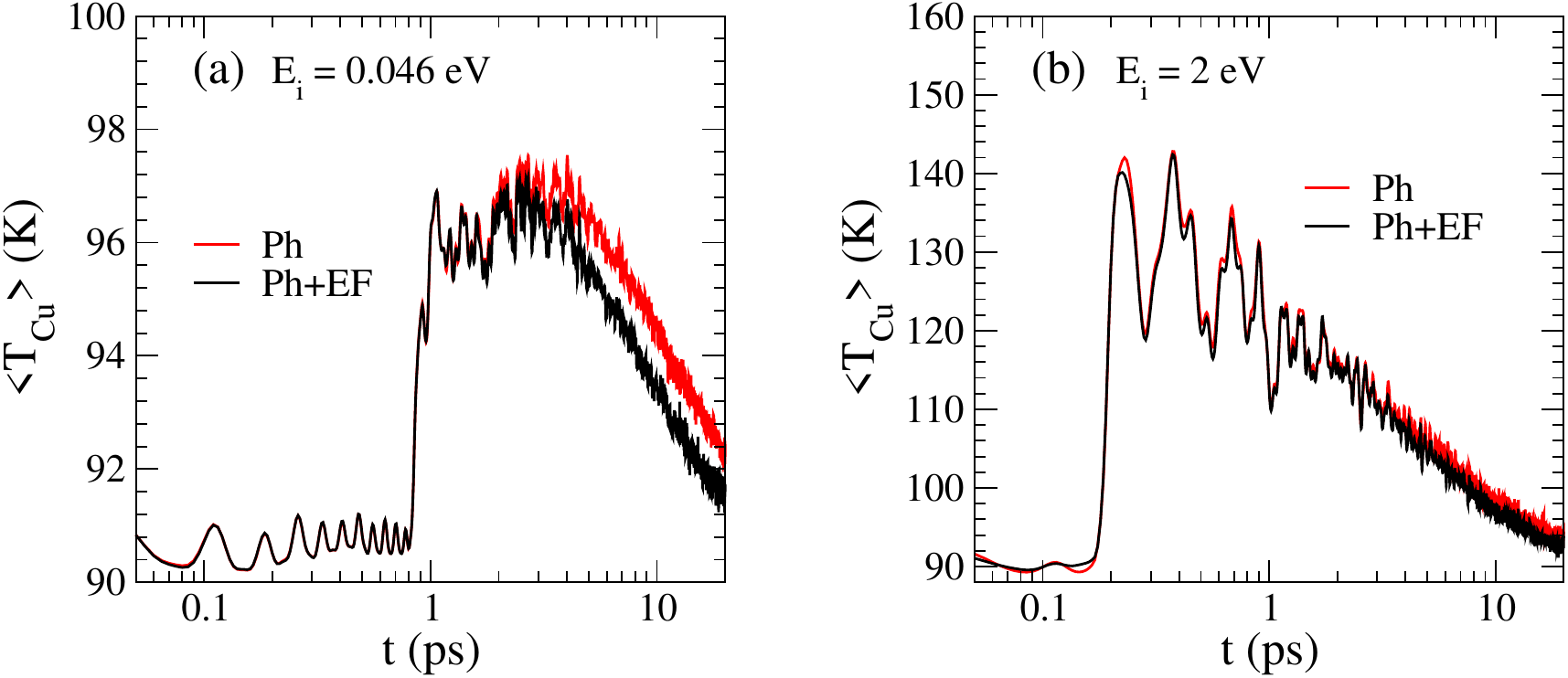} 
	\caption{Time evolution of the mean temperature of mobile Cu atoms, $\left\langle T_{\text{Cu}} \right\rangle $, for adsorbed CO molecules: (a) E$_{\mathrm{i}}$=0.046 eV, (b) E$_{\mathrm{i}}$=2 eV.
    Red and black curves correspond to Ph and Ph+EF simulations, respectively.}   
    \label{fig:Temp_6x5}
\end{figure}

In addition, Fig.~\ref{fig:Temp_6x5} shows the time evolution of the mean temperature of mobile Cu atoms, $\left\langle T_\text{Cu} \right\rangle = 2\left\langle E_\text{kin}^{Cu}\right\rangle/(3\,k_B\,N_{Cu})$. 
Here, $\left\langle E_\text{kin}^{Cu}\right\rangle$ is the ensemble average of the kinetic energy for the $N_{Cu}=90$ mobile Cu atoms, calculated across the set of trajectories classified as adsorbed. Data are shown for E$_i$=0.046 eV and 2 eV, which are representative of the low- and high-energy regimes, respectively.

For all the incidence energies considered (0.046, 0.575, 1.2, and 2 eV), $\left\langle E_{kin}\right\rangle$ exhibits an initial time interval from 0 to $t_1$ (where $t_1$ corresponds to the time of the first minimum in $\left\langle Z_{cm}\right\rangle$, the average center-of-mass distance from the surface), during which the molecules first experience an acceleration towards the surface due to the long-range attractive potential, as analyzed below.

Between $t_1$ and $t_2$ (where $t_2=1.5$ ps), the $\left\langle E_{kin}\right\rangle$ curves exhibit oscillatory behavior. These oscillations correspond to the molecules undergoing multiple brief bouncing events on the surface. This transition time interval is short, and its duration increases with E$_i$. 

For $t\geq t_2$, the $\left\langle E_{kin}\right\rangle$ curves decrease monotonically (following an approximately exponential relationship according to the semi-log scale), as the molecules lose energy to the surface. 
Also, the molecules remain near the surface 
$\left\langle Z_{cm} \right\rangle<3$ \AA,  indicating that the molecules are already, on average, in a zone where the electronic density generated by the surface atoms is high.
The rate of decrease for the Ph+EF simulations is higher than the rate observed for Ph simulations for $t\gtrsim  t_2$ ps,  which suggests that the long duration of interaction with the molecule close to the surface (where the electron density is high) causes the electronic friction effect to become significant.

Consequently, the addition of electronic friction in the Ph+EF model causes a more rapid dissipation of the CO kinetic energy during accommodation into the chemisorption well. Note that, although $\left\langle E_{kin}\right\rangle$  curves for the two models begin to separate at $t\approx t_2$ ps, the equilibrium height of $\left\langle Z_{cm} \right\rangle$ ($\approx$2.65 \AA) is not reached until times above 4 or 5 ps. This strongly connects with the previously defined adsorption time interval from 0 to 5 ps, confirming that the greater effects of e-h pair excitations occur during the final phase of accommodation.

The faster efficient energy transfer to phonons compared to electrons is illustrated in Fig.~\ref{fig:Temp_6x5}. For both impact energies considered, $\left\langle T_\text{Cu} \right\rangle $ exhibits a sharp increase during the first picosecond, this reflecting the rapid transfer of momentum from the impinging molecule to the surface atoms. The temperature rise is more pronounced --and occurs at earlier times-- for E$_{\mathrm{i}}$=2 eV, consistent with the larger kinetic energy available of the impinging molecule.  
At longer times, $\left\langle T_\text{Cu} \right\rangle $ gradually returns toward the initial value T$_{\mathrm{s}}$=90 K.

As shown in Figs. \ref{fig:Temp_6x5}(a) and (b), the time evolution of $\left\langle T_\text{Cu} \right\rangle $ is very similar for both Ph and Ph+EF simulations. A slightly more rapid cooling is observed when electronic friction is included, which is more apparent in the low energy case. When electronic friction is included, the energy of the CO molecule is lower and, therefore, the molecule transfers (on average) less energy in its collisions with the surface atoms. As a consequence, thermal equilibration is slightly faster.

The overall results from Figs.~\ref{fig:Ekintot_6x5} and \ref{fig:Temp_6x5} indicate that most of the total kinetic energy of the molecule is transferred to the Cu lattice within the first picoseconds of the simulation, confirming that phonon-mediated dissipation dominates the initial adsorption stage. Furthermore, the present results are in line with previous AIMDEF studies on other systems, which showed that phonon-mediated dissipation dominates the early adsorption stages of N and N$_{2}$ on Pd(100), Ag(111), and Fe(110), whereas electronic excitations remain active over longer time scales \cite{Novko2015,Novko2016a,Novko2016b,Alducin2017}.

\begin{figure}[htb!]
	\includegraphics[width=0.7\textwidth]{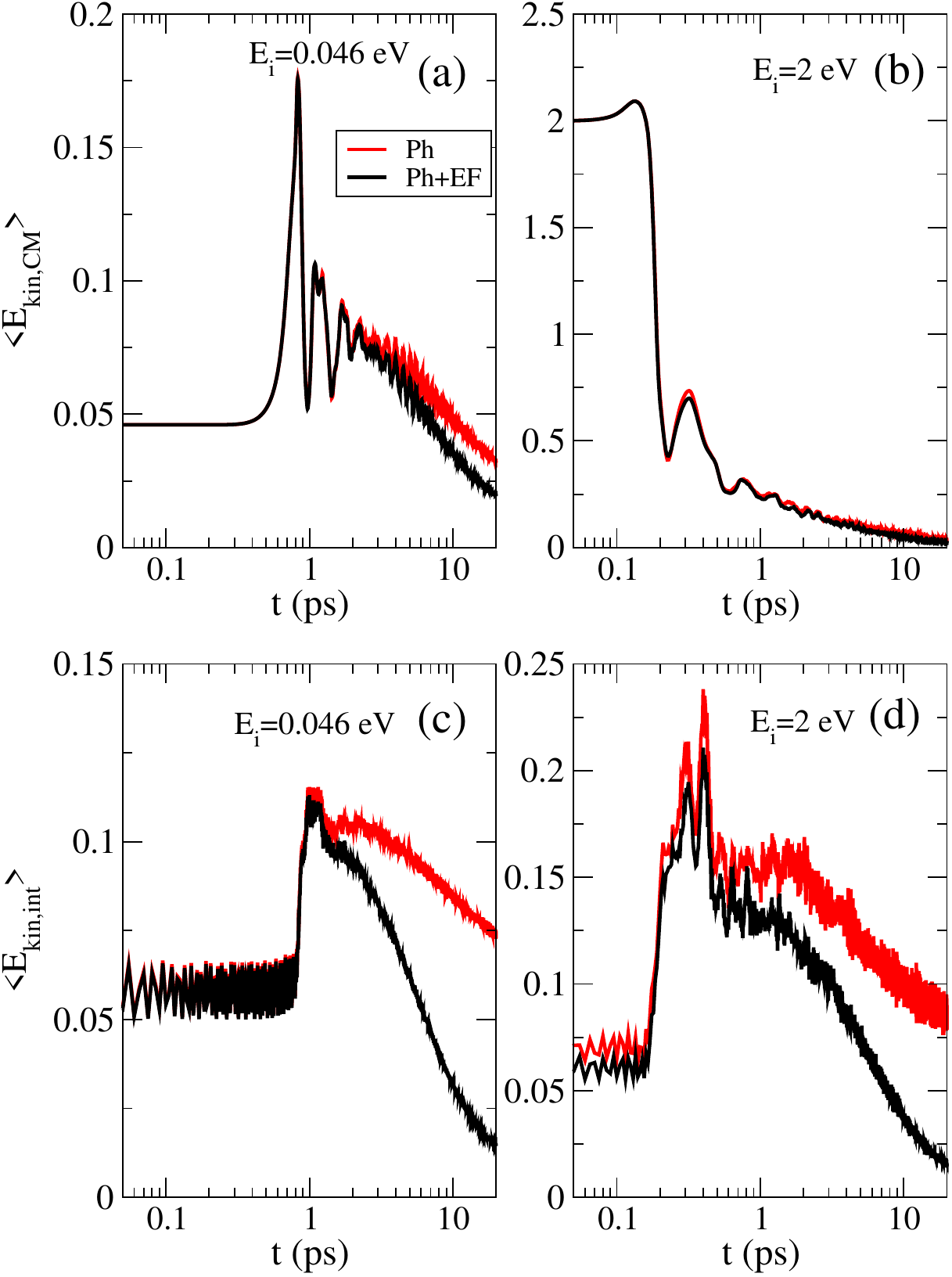}   
	\caption{Upper panels: $\left\langle E_{kin,CM}\right\rangle$ values at different incidence energies: (a) 0.046 eV, (b) 2.0 eV. Lower panels:
	$\left\langle E_{kin,int}\right\rangle $ values at different incidence energies: (c) 0.046 eV, (d) 2.0 eV (d). 
    Red and black curves correspond to Ph and Ph+EF simulations, respectively.}   
    \label{fig:EkinCM_6x5}
\end{figure}

It is interesting to investigate in more detail the energy exchange of the different degrees of freedom of the adsorbate with the surface atoms as predicted by our model, particularly to disentangle the effect of including electronic friction. 
To this aim, Fig.~\ref{fig:EkinCM_6x5} displays the mean kinetic energies of the molecular center-of-mass, $\left\langle E_{kin,CM} \right\rangle$, and of the internal degrees of freedom, $\left\langle E_{kin,int} \right\rangle$, where $E_{kin,int}=E_{kin}-E_{kin,CM}$, and the initial conditions are  $\left\langle E_{kin,int} (0) \right\rangle=$ZPE/2 and $\left\langle E_{kin,CM} (0) \right\rangle=$E$_i$.

The evolution of $\left\langle E_{kin,CM} \right\rangle $ in Figs. \ref{fig:EkinCM_6x5}(a)  and (b) closely follows the total kinetic energy trend: a sharp initial peak followed by rapid decay. The Ph and Ph+EF curves remain remarkably similar because the translational energy is drastically reduced via lattice vibrations during the adsorption time interval ($t\le 5$ ps). Both at high (E$_i= 2 $ eV) and low (E$_i= 0.046 $ eV) incidence energy, the majority of the translational kinetic energy is transferred to the Cu atoms upon the first impact. Consequently, by the time electronic friction becomes relevant, the remaining CM velocity is already low, making the additional dissipation into the electronic channel minor in absolute terms.

In contrast, the internal kinetic energy,  $\left\langle E_{kin,int} \right\rangle $, shown in Figs. \ref{fig:EkinCM_6x5}(a) and (b), reveals greater differences between the two models. In Ph simulations, the molecule retains significant internal energy because high-frequency internal modes couple inefficiently to the surface phonons which presumably have lower-frequency modes. However, the Ph+EF simulations show a rapid decay, eventually driving the internal energy below the molecular ZPE/2 value. While this relaxation is a known artifact of classical calculations of CO molecule toward thermal equilibrium ($k_B \text{T}_{ph}\approx 7.75$ meV $\ll$ ZPE/2), it qualitatively demonstrates that e-h pair excitations are considerably important for damping internal degrees of freedom where phonons are ineffective.

The observed difference in how electronic excitations affect translational CM versus internal molecular kinetic energy for $t \gtrsim t_2$ does not necessarily 
mean that e-h pair generation promotes a stronger coupling with internal molecular degrees of freedom. 
Instead, these results suggest that the extra energy dissipation from electronic friction only becomes significant once the molecule has lost most of its initial translational CM kinetic energy and thus lacks sufficient energy to continue moving extensively on the surface. 
This is further supported by the low values observed for $\left\langle Z_{cm} \right\rangle$ ($\left\langle Z_{cm} \right\rangle<3$ \AA). This suggests that e-h pairs become relevant primarily during the final stages of thermalization (accommodation period), after the molecule has been adsorbed onto the surface. 
To confirm this, we require a detailed analysis of the final state of adsorbed molecules, which is provided in the following section.

\subsection{Final state of adsorbed molecules}
\label{sec:characterization}

\begin{figure}[htb!]
	\includegraphics[width=0.75\linewidth]{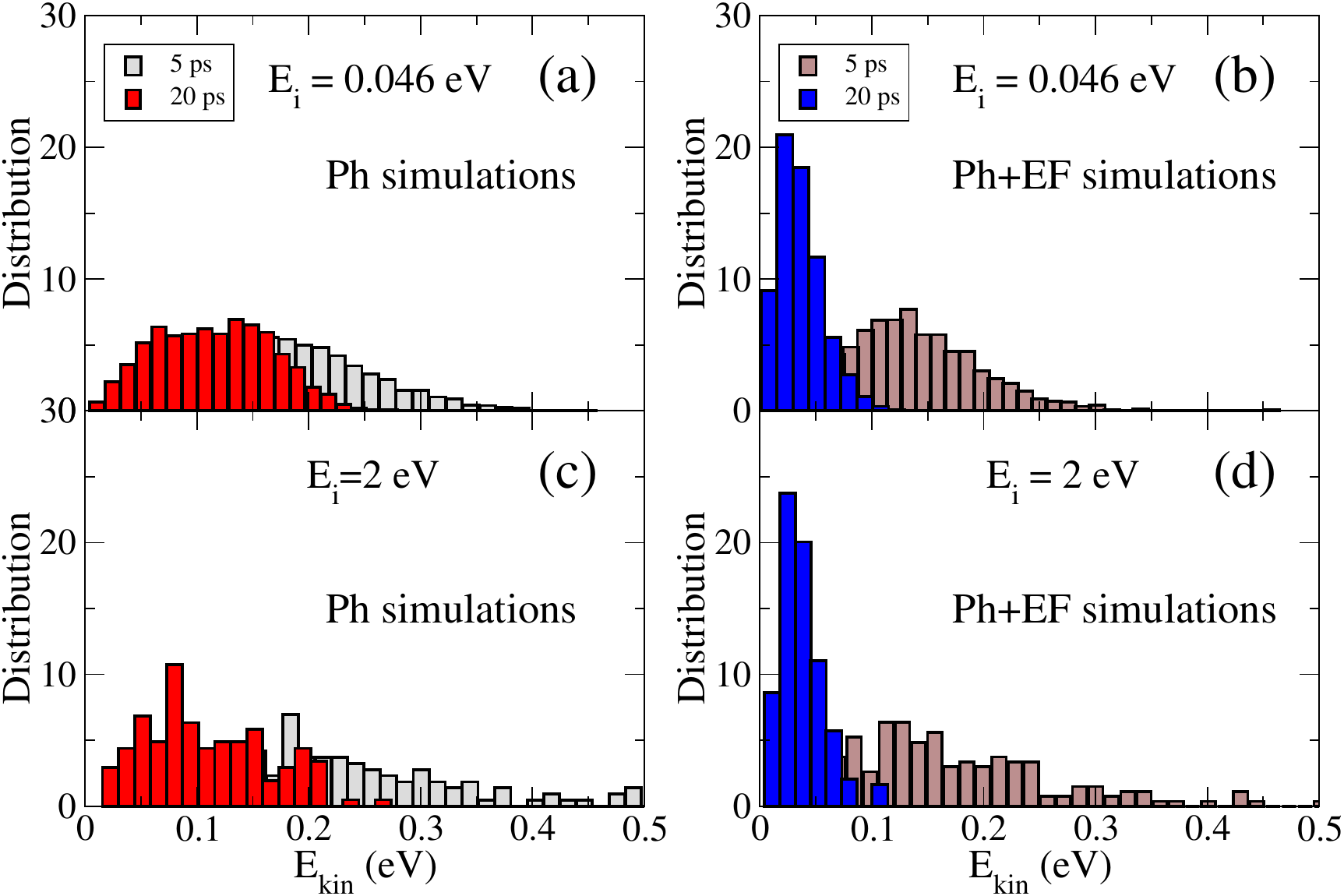}
	\caption{\label{fig:EkinHisto_6x5}Distributions of the total kinetic energy of adsorbed CO molecules, $E_{kin}$, at 5 and 20 ps. (a) Ph and (b) Ph+EF simulations for E$_{\mathrm{i}}$=0.046 eV. (c) Ph and (d) Ph+EF simulations for E$_{\mathrm{i}}$=2 eV. Gray and brown (red and blue) histograms correspond to 5 ps (20 ps).}   
\end{figure}

In order to characterize the final state of adsorbed CO molecules, we show in Fig.~\ref{fig:EkinHisto_6x5} the distributions of total kinetic energies for Ph and Ph+EF simulations at 5 and 20 ps for two incidence energies: E$_i$=0.046 and E$_i$=2 eV. These distributions are normalized by unity area.

For both incidence energies, the distributions narrow and shift toward lower kinetic energies as time increases, reflecting the progressive energy loss of the adsorbate. The histograms for Ph and Ph+EF are rather similar at 5 ps. In contrast, the results at 20 ps show that the inclusion of electronic friction produces a more pronounced shift of the peak toward lower energies. 

\begin{figure}[htb!]
	\includegraphics[width=0.9\linewidth,trim={0cm 0cm 0cm 0cm}]{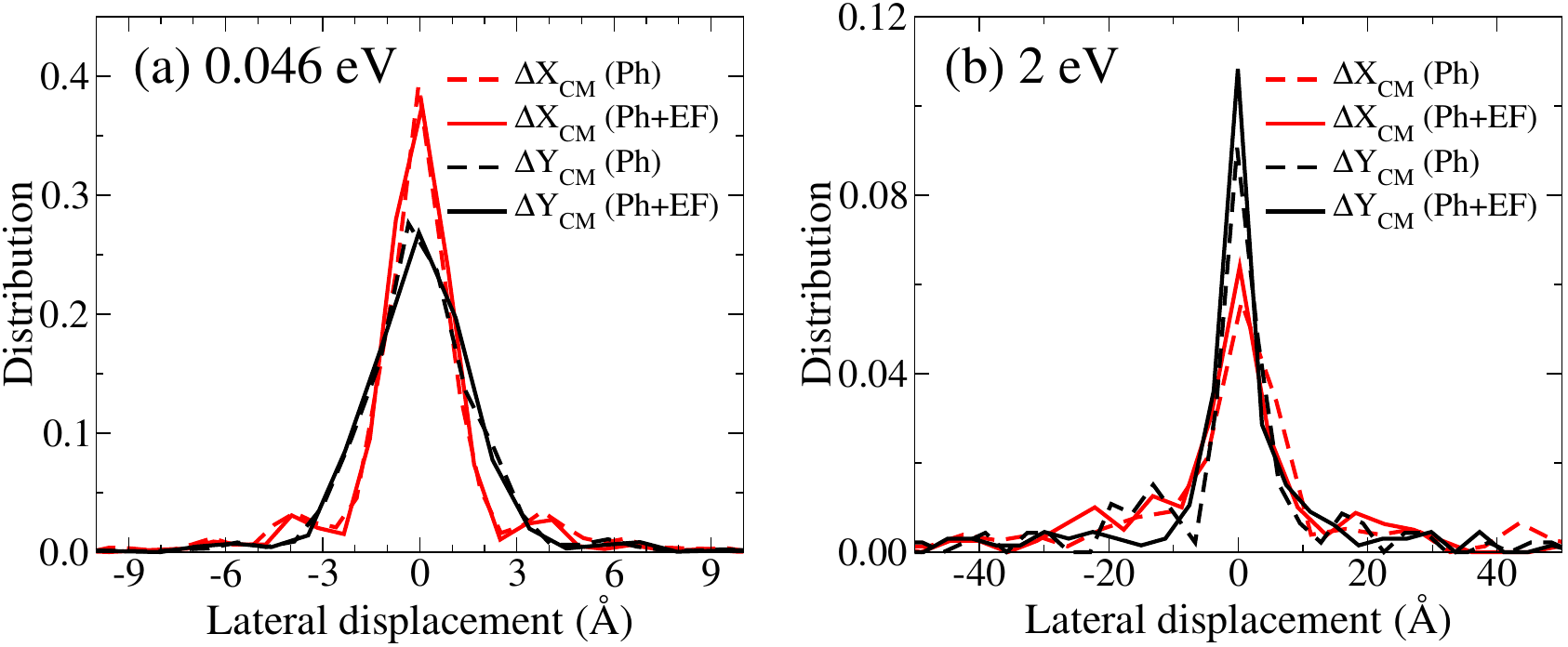} 
	\caption{Distributions of lateral displacements of the CO center of mass along the $X$ and $Y$ directions at $\mathrm{t_{f}}$=20 ps for Ph and Ph+EF simulations. (a) E$_{\mathrm{i}}$=0.046 eV. (b) E$_{\mathrm{i}}$=2 eV.}  
    \label{fig:Disp_6x5}
\end{figure}

This observed difference raises the question of whether the molecule exhibits reduced lateral motion at 20 ps when electronic friction is included. To address this question,  Fig. \ref{fig:Disp_6x5} shows the distributions (normalized to unity area) of lateral displacements of the molecular center of mass for adsorbed trajectories, ($\Delta X_{CM}$, $\Delta Y_{CM}$), defined as $\Delta X_{CM}=X_{CM}(20\,ps)-X_{CM}(0\,ps)$ and $\Delta Y_{CM}=Y_{CM}(20\,ps)-Y_{CM}(0\,ps)$.
It can be seen that the incorporation of electronic friction (solid lines compared to dashed lines) does not appreciably modify the distributions of lateral displacements for both directions and both analyzed energies. 
This confirms that the inclusion of electronic friction does not influence the spatial localization of the trapped molecules in this system.

Regarding the specific dynamics at different energies, for E$_{\mathrm{i}}$=0.046 eV, the distributions extend up to $\approx5$ \AA~ in both directions, a range related to the unit cell dimensions. The nearest-neighbor distance between Cu atoms along the $X$ direction is 2.65 \AA, which corresponds to the observed width of $\Delta X_{CM}$, while along the $Y$ direction it is 3.754 \AA, which corresponds to the observed width of $\Delta Y_{CM}$. This shows that the CO molecules do not move laterally more than one unit cell with respect to their initial positions.

In contrast, for the higher incidence energy, E$_{\mathrm{i}}$=2 eV, the molecules retain significant kinetic energy, allowing them to reach larger lateral distances. The distributions of $\Delta X_{CM}$ are slightly broader than those of $\Delta Y_{CM}$ (as observed by the higher peak values for the $\Delta Y_{CM}$ distributions), indicating that lateral motion is favored along the $\left[ \bar{1}10 \right] $ ($X$) direction. This higher mobility along $X$ is consistent with the potential energy landscape\cite{Gonzalez2023,Gonzalez2025}, where the activation energy for diffusion between neighboring top sites along $X$ is low ($\approx$0.13 eV) compared to the higher activation energy ($\approx$0.49 eV) along the $Y$ direction, which makes lateral movement more restricted in the latter case.

\begin{figure}[htb!]
	\begin{subfigure}[b]{\linewidth}
		\includegraphics[width=0.8\linewidth]{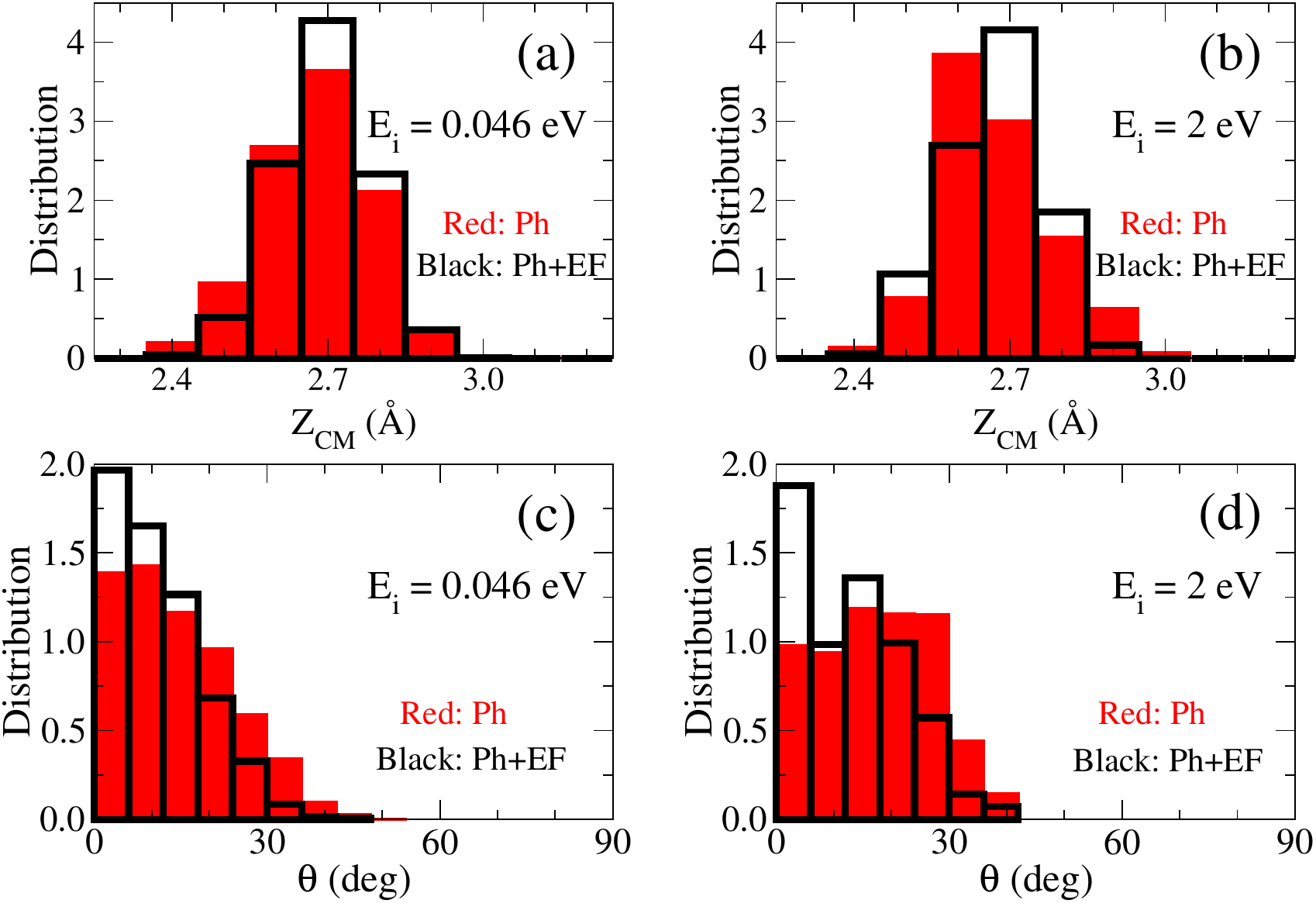}
	\end{subfigure}
	\\
	\begin{subfigure}[b]{\linewidth}
		\includegraphics[width=0.3\linewidth,trim={0cm 0cm 0cm 0cm}]{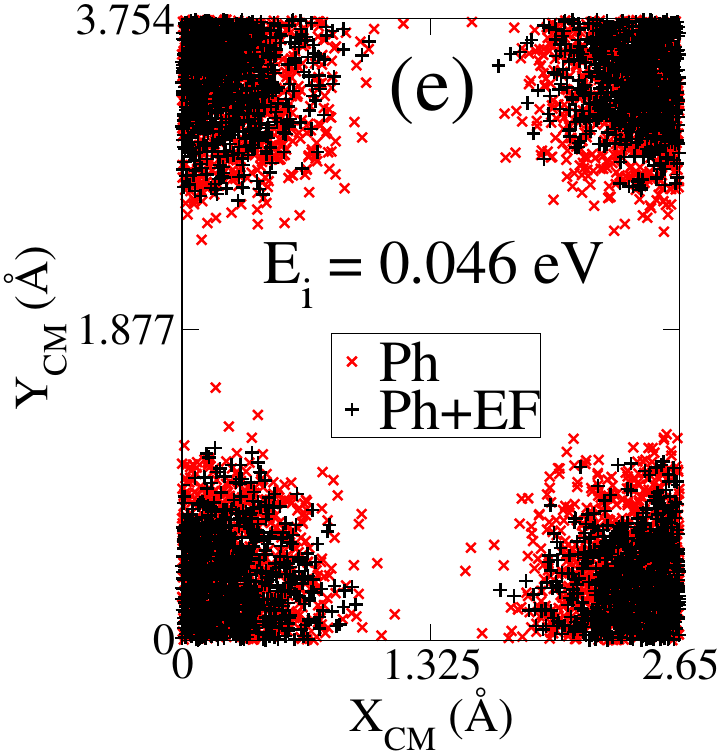}
	\end{subfigure}
	\caption{Geometric characterization of chemisorbed CO molecules at 20 ps. Panels (a) and (b): distributions for the $Z_{CM}$ coordinate at: (a) E$_{\mathrm{i}}$=0.046 eV, and (b) E$_{\mathrm{i}}$=2 eV. Panels (c) and (d): distributions for the  orientation polar angle $\theta$ at: (c) E$_{\mathrm{i}}$=0.046 eV, (d) E$_{\mathrm{i}}$=2 eV. (e) Distribution for the lateral position of the molecular CM, ($X_{CM}$,$Y_{CM}$) at E$_{\mathrm{i}}$=0.046 eV. Red and black colors correspond to Ph and Ph+EF simulations, respectively.}    \label{fig:geometric}
\end{figure}

The adsorption geometry of CO molecules trapped at $\mathrm{t_{f}}$=20 ps is summarized in Fig. \ref{fig:geometric}. 
The distributions of $Z$ coordinates (normalized by unity area) of the molecular center of mass ($Z_{CM}$), shown in panels (a) and (b), are narrowly centered around $Z_{CM}\approx$ 2.65 {\AA}, which corresponds to the chemisorption minimum previously identified in our vdW-DF2 calculations \cite{Gonzalez2023}. The inclusion of electronic friction slightly modifies the shape of the distributions, producing a marginally narrower peak (as can be seen by the higher value at $Z_{CM}\approx$ 2.65 {\AA} for E$_i$=2 eV.   The results for E$_i$=2 eV exhibit higher discrepancies due to the increased statistical uncertainty associated with fewer adsorbed trajectories at this incidence energy.

The molecular polar angle ($\theta$) distributions, divided by $\sin(\theta)$ and normalized to unity area, are shown in Fig.~\ref{fig:geometric} (c) and (d), These panels reveal that the orientation of the chemisorbed CO molecules at 20 ps is predominantly perpendicular to the surface ($\theta\approx 0$ deg.) with the C atom pointing toward the surface. 
The inclusion of electronic friction induces a small narrowing toward lower $\theta$ values, suggesting a weak tendency of the molecule to adopt a more perpendicular alignment as a result of enhanced energy dissipation.

Regarding the 2D-distributions of the lateral center of mass coordinates ($X_{CM}$, $Y_{CM}$) shown in Fig.~\ref{fig:geometric} (e), only the results for E$_{\mathrm{i}}$=0.046 eV are presented, as the number of adsorbed trajectories for E$_i$=2 eV is statistically insufficient to generate a meaningful visualization. 
The Ph (red) and Ph+EF (black) schemes exhibit a well-defined zone around the top sites of the Cu(110) lattice, in agreement with the stable chemisorption geometry characterized in our previous work\cite{Gonzalez2023}. 
While the overall 2D-distributions are very similar, a subtle visual inspection of the regions with a higher density of points suggests a slight narrowing when electronic friction is included. This slight change could be interpreted as a small tendency for the molecule to be more localized toward the expected equilibrium position. This observation is consistent with previous discussions of faster energy dissipation when electronic friction is included, although the effect on the final localization remains minimal.

\section{Conclusions}
In this work, we have investigated the relative roles of phononic and electronic energy dissipation in CO adsorption on Cu(110) using classical trajectory simulations based on a previously developed and validated vdW-DF2 ANN-PES \cite{Gonzalez2023}. Initial sticking probabilities (S$_{0}$) were calculated as a function of impact energies (E$_{\mathrm{i}}$) for a surface initially thermalized at T$_{\mathrm{S}}$=90 K. 
We compared the obtained S$_0$ values for two supercell sizes (3$\times$2) and (6$\times$5), showing that the better description of the finite-size effects for the (6$\times$5) supercell yields to slightly higher S$_0$ values and thus, a better agreement with available experimental data. This consistency confirms that the present theoretical approach accurately captures the essential multidimensional energy landscape required to describe the system. We also compared the obtained S$_0$ values for simulations with and without nonadiabatic effects (incorporated with the electronic friction approach under LDFA), 
showing that the inclusion of electronic friction produces, at most, a marginal increase in the S$_0$ values.

A crucial finding of our analysis is that the molecular adsorption of CO is practically determined within the first 5 ps of simulation. We observed that molecules remaining trapped near the surface after this time can be considered adsorbed with negligible error, as the deviations in computed sticking probabilities and reflection events between 5 ps and 20 ps are minimal for all the impact energies studied. Furthermore, the analysis of energy dissipation reveals distinct behaviors before and after this time threshold. During the first 5 ps, the decay of the molecular mean kinetic energy is nearly identical, regardless of whether electronic friction is included or not. This demonstrates that the excitation of surface phonons is the overwhelmingly dominant mechanism for the initial energy loss and the subsequent trapping of the molecule. At longer times (t>5 ps), however, the influence of the electronic channel becomes apparent. The inclusion of electronic friction introduces a smooth and systematic additional damping to the energy relaxation process. While this late-stage dissipation into electron-hole pair excitations does not alter the final sticking probabilities, it becomes a significant pathway for dissipating the remaining energy, particularly at low impact energies where the coupling of internal molecular degrees of freedom to the electronic system becomes increasingly important.

Regarding the final state of the adsorbed molecules, we also examined the influence of nonadiabatic effects on the lateral motion,  orientation, and position at 20 ps. No significant changes in the lateral motion were observed when comparing results with and without electronic friction. 
This confirms that the additional energy dissipation provided by the electronic channel (from 5 to 20 ps) does not significantly alter the adsorption site at which the molecule is located at 20 ps. 
Consequently, the molecules stabilize in the expected chemisorption geometry, centered at the top positions of the Cu(110) lattice. The inclusion of electronic friction on the adsorption site selection is negligible, though it leads to a marginal narrowing of the kinetic energy distributions and an increased rate of stabilization toward equilibrium geometry.

In summary, our results establish a clear hierarchy of dissipation mechanisms for CO on Cu(110). Phonon-mediated dissipation is the governing factor for the initial sticking probability, efficiently removing the necessary energy for trapping within the first few picoseconds. In contrast, electron-hole pair excitations play a secondary role, contributing negligibly to the trapping probability but becoming relevant for accelerating the subsequent thermalization of the adsorbate. This distinction implies that for systems characterized by moderate chemisorption wells ($\approx$0.6 eV) and relatively heavy adsorbates, such as CO/Cu(110), accurate predictions of sticking probabilities can be achieved with phonon-only models, while a complete description of the long-term energy relaxation dynamics requires the inclusion of nonadiabatic effects.

\begin{acknowledgments}
C.A.T., F.J.G., and H.F.B. acknowledge financial support from the ANPCyT Project No.~PICT-2021-I-A-01135, CONICET Project No.~PIP 1679, and UNR Project No.~PID 80020190100011UR (Argentina). A.P.S., J.I.J, and M.A. acknowledge financial support provided by the Spanish MCIN$/$AEI$/$10.13039$/$501100011033$/$, FEDER Una manera de hacer Europa (Grant No.~PID2022-140163NB-I00), Gobierno Vasco-UPV/EHU (Project No.~IT1569-22), and the Basque Government Education Departments' IKUR program, also co-funded by the European NextGenerationEU action through the Spanish Plan de Recuperación, Transformación y Resiliencia (PRTR). Authors acknowledge the computer time provided by CCT-Rosario Computational Center, member of the High Performance Computing National System of Argentina (SNCAD)
\end{acknowledgments}

\section*{Data Availability Statement}
The data that support the findings of this study are available from the corresponding author upon reasonable request.






%
\bibliographystyle{aipnum4-1}
\bibliography{References}




\end{document}